\documentclass{article}
\usepackage{amsmath}
\usepackage{cite}
\usepackage{graphicx}
\usepackage{dcolumn}

\begin{document}

\date{}
\title{Two simple models derived from a quantum-mechanical particle on an
elliptical path}
\author{Francisco M. Fern\'{a}ndez\thanks{%
fernande@quimica.unlp.edu.ar} \\
INIFTA, DQT, Sucursal 4, C. C. 16, \\
1900 La Plata, Argentina}
\maketitle

\begin{abstract}
We analyze two simple models derived from a quantum-mechanical particle on
an elliptical path. The first Hamiltonian operator is non-Hermitian but
equivalent to an Hermitian operator. It exhibits the same two-fold
degeneracy as the particle on a circular path. More precisely, the spectrum
is $E_{n}=n^{2}E_{1},\ n=0,\pm 1,\pm 2,\ldots $, $E_{1}>0$. The second
Hamiltonian operator is Hermitian and does not exhibit such degeneracy. In
this case the nth excited energy level splits at the nth order of
perturbation theory. Both models can be described in terms of symmetry point
groups with one-dimensional irreducible representations but the
non-Hermitian example exhibits an additional symmetry that explains the
two-fold degeneracy. The Schr\"{o}dinger equation for the first example is
exactly solvable. \newline
\end{abstract}

\section{Introduction}

\label{sec:intro}

Most textbooks on quantum mechanics and quantum chemistry\cite{P68} resort
to exactly-solvable models in order to illustrate some of the features of
quantum-mechanical systems. Among the simplest models one finds the particle
in a box, the harmonic oscillator and the planar rigid rotator. The latter
model is one of the few cases of one degree of freedom that exhibits
degeneracy. It is mathematically similar to a single particle moving along a
circular path. An interesting deformation of this model is the case of a
particle moving along an ellipse. The purpose of this paper is a
straightforward analysis of the latter model.

In section~\ref{sec:model} we derive the Hamiltonian operator for the model
and discuss the possible scalar product for the states as well as other
useful features. In section~\ref{sec:PT} we derive an analytical expression
for the spectrum from first-order perturbation theory. In section~\ref
{sec:RRM} we obtain accurate eigenvalues by means of the Rayleigh-Ritz
method (RRM)\cite{P68} that yields increasingly accurate upper bounds\cite
{M33, F25a}. From the secular determinant we derive perturbation corrections
of greater order. In section~\ref{sec:alternative_mod} we introduce a
slightly different alternative model and carry out similar calculations.
Finally, in section~\ref{sec:conclusions} we summarize the main results of
the paper and present our conclusions.

\section{The model}

\label{sec:model}

We consider a particle of mass $m$ that moves freely on an elliptical path.
The Hamiltonian operator is
\begin{equation}
H=-\frac{\hbar ^{2}}{2m}\nabla ^{2},  \label{eq:H}
\end{equation}
where $\nabla ^{2}$ is the Laplacian in two dimensions. The motion of the
particle is restricted to a closed path given by all points $(x,y)$ that
satisfy
\begin{equation}
\frac{x^{2}}{a^{2}}+\frac{y^{2}}{b^{2}}=1,  \label{eq:ellipse}
\end{equation}
where $a$ and $b$ are positive real numbers (the ellipse semi-axes). If we
write
\begin{equation}
x=a\cos \phi ,\;y=b\sin \phi ,  \label{eq:x,y(phi)}
\end{equation}
which satisfy equation (\ref{eq:ellipse}), we obtain
\begin{eqnarray}
\nabla ^{2} &=&\frac{1}{\sqrt{a^{2}+\left( b^{2}-a^{2}\right) \cos ^{2}\phi }%
}\frac{d}{d\phi }\frac{1}{\sqrt{a^{2}+\left( b^{2}-a^{2}\right) \cos
^{2}\phi }}\frac{d}{d\phi }=\frac{1}{a^{2}}\tilde{\nabla}^{2},  \nonumber \\
\tilde{\nabla}^{2} &=&\frac{1}{\sqrt{1+\xi \cos ^{2}\phi }}\frac{d}{d\phi }%
\frac{1}{\sqrt{1+\xi \cos ^{2}\phi }}\frac{d}{d\phi },\;\xi =\frac{%
b^{2}-a^{2}}{a^{2}},  \label{eq:Laplacian_phi}
\end{eqnarray}
where $-1<\xi <\infty $. The point $\xi =-1$ is expected to be a singularity
because it leads to $b=0$.

We can define a dimensionless Hamiltonian operator as\cite{F20}
\begin{equation}
\tilde{H}=\frac{2ma^{2}}{\hbar ^{2}}H=-\tilde{\nabla}^{2},  \label{eq:H_dim}
\end{equation}
and from now on we will focus on the dimensionless eigenvalue equation
\begin{eqnarray}
H\psi _{n}(\phi ) &=&E_{n}\psi _{n}(\phi ),\;\psi _{n}(\phi +2\pi )=\psi
_{n}(\phi ),  \nonumber \\
H &=&-A^{2},\;A=\frac{1}{\sqrt{g}}\frac{d}{d\phi },\;g=1+\xi \cos ^{2}\phi .
\label{eq:Schro}
\end{eqnarray}

The Hamiltonian operator (\ref{eq:Schro}) is Hermitian with respect to the
scalar product
\begin{equation}
\left\langle u\right| \left. v\right\rangle =\int_{0}^{2\pi }u^{*}(\phi
)v(\phi )\sqrt{g}\,d\phi ,
\end{equation}
where $\sqrt{g}$ is the Jacobian of the transformation. However, this scalar
product exhibits two difficulties. The first one is that the only variable
parameter $\xi $ of the dimensionless model appears in it. The second
drawback is that it makes the calculation of matrix elements more
complicated. For these reasons, in what follows we choose
\begin{equation}
\left\langle u\right| \left. v\right\rangle =\int_{0}^{2\pi }u^{*}(\phi
)v(\phi )\,d\phi ,  \label{eq:scalar_prod}
\end{equation}
which facilitates the numerical calculation based on Fourier basis sets
(although it somewhat changes the nature of the model). As a result, the
Hamiltonian operator $H$ is not Hermitian because
\begin{equation}
H^{\dagger }=-\frac{d}{d\phi }\frac{1}{\sqrt{g}}\frac{d}{d\phi }\frac{1}{%
\sqrt{g}}\neq H.  \label{eq:H_adjoint}
\end{equation}

However, it follows from
\begin{equation}
g^{1/4}Hg^{-1/4}=g^{-1/4}H^{\dagger }g^{1/4}=\mathcal{H}=-g^{-1/4}\frac{d}{%
d\phi }g^{-1/2}\frac{d}{d\phi }g^{-1/4},  \label{eq:isomorphism}
\end{equation}
that both $H$ and $H^{\dagger }$ are equivalent to the Hermitian operator $%
\mathcal{H}$ and, therefore, share the same real spectrum. Non-Hermitian
operators with this property have been discussed since long ago\cite{P43,
W69, SGH92} (a review of related papers is given elsewhere\cite{F22}).

If $H\psi =E\psi $ and $H^{\dagger }\varphi =E\varphi $ then one can easily
prove that the Hellmann-Feynman theorem (HFT)\cite{G32, F39} in this case
reads\cite{HMM24}
\begin{equation}
\frac{dE}{d\xi }=\frac{\left\langle \varphi \right| \frac{dH}{d\xi }\left|
\psi \right\rangle }{\left\langle \varphi \right| \left. \psi \right\rangle }%
.  \label{eq:HFT}
\end{equation}
The HFT is valid even for degenerate states as discussed elsewhere for
Hermitian operators\cite{F24}.

The Hamiltonian operator $H$ is invariant under the operations $\phi
\rightarrow -\phi $ and $\phi \rightarrow \phi +\pi $; therefore, we expect
that $\psi (-\phi )=\pm \psi (\phi )$ and $\psi (\phi +\pi )=\pm \psi (\phi
) $. For this reason we can separate the states into four classes given by $%
(+,+)$, $(+,-)$, $(-,+)$ and $(-,-)$ from which we conclude that the problem
can be described by means of the symmetry point groups $D_{2}$ or $C_{2v}$,
both with one-dimensional irreducible representations\cite{C90}. It is clear
that the functions belonging to the classes $(+,+)$ and $(-,+)$ are periodic
of period $\pi $. The relation between present notation and the one in
standard books on group theory\cite{C90} is given in the following table
\begin{equation}
\begin{array}{ccc}
D_{2} & C_{2v} & \mathrm{\Pr esent} \\
A & A_{1} & (+,+) \\
B_{1} & A_{2} & (-,+) \\
B_{2} & B_{1} & (+,-) \\
B_{3} & B_{2} & (-,-)
\end{array}
.  \label{eq:Group_table}
\end{equation}

\section{Perturbation theory}

\label{sec:PT}

When $\xi =0$ the dimensionless Hamiltonian operator becomes $H_{0}=-\frac{%
d^{2}}{d\phi ^{2}}$ so that
\begin{equation}
H_{0}\psi _{n}^{(0)}=E_{n}^{(0)}\psi _{n}^{(0)},\;E_{n}^{(0)}=n^{2},\;\psi
_{n}^{(0)}(\phi )=\frac{1}{\sqrt{2\pi }}e^{in\phi },\;n=0,\pm 1,\pm 2,\ldots
.  \label{eq:Schro_H_0}
\end{equation}
By means of perturbation theory (PT) we can obtain approximate solutions in
terms of power series
\begin{equation}
E_{n}=\sum_{j=0}^{\infty }E_{n}^{(j)}\xi ^{j},\;\psi _{n}=\sum_{j=0}^{\infty
}\psi _{n}^{(j)}\xi ^{j}.  \label{eq:PT series}
\end{equation}
Note that there is always an exact solution given by $E_{0}=E_{0}^{(0)}=0$
and $\psi _{0}(\phi )=\psi _{0}^{(0)}(\phi )$ for all $\xi $.

The perturbation correction of first order can be derived by means the
non-Hermitian operator
\begin{equation}
H_{1}=\left. \frac{dH}{d\xi }\right| _{\xi =0}=\cos ^{2}\phi \frac{d^{2}}{%
d\phi ^{2}}-\sin (\phi )\cos (\phi )\frac{d}{d\phi }.  \label{eq:H_1}
\end{equation}
Since
\begin{equation}
\left\langle \psi _{-n}^{(0)}\right| H_{1}\left| \psi
_{n}^{(0)}\right\rangle =\left\langle \psi _{n}^{(0)}\right| H_{1}\left|
\psi _{-n}^{(0)}\right\rangle =0,\;\left\langle \psi _{n}^{(0)}\right|
H_{1}\left| \psi _{n}^{(0)}\right\rangle =\left\langle \psi
_{-n}^{(0)}\right| H_{1}\left| \psi _{-n}^{(0)}\right\rangle =-\frac{n^{2}}{2%
},  \label{eq:H_1_mat_el}
\end{equation}
we conclude that
\begin{equation}
E_{n}=n^{2}\left( 1-\frac{\xi }{2}\right) +\mathbf{O}\left( \xi ^{2}\right) .
\label{eq:E_n_PT}
\end{equation}
We obtain exactly the same result using $H_{1}^{\dagger }$ as expected from
the argument given in the preceding section. Besides, the HFT at $\xi =0$%
\begin{equation}
\left. \frac{dE}{d\xi }\right| _{\xi =0}=-\frac{n^{2}}{2},  \label{eq:HFT2}
\end{equation}
predicts that all the eigenvalues have a negative slope at origin.

Since the eigenvalues are expected to be singular when $\xi =-1$ it appears
convenient to try the improved perturbation approximation
\begin{equation}
E_{n}\approx \left\langle \psi _{n}^{(0)}\right| H\left| \psi
_{n}^{(0)}\right\rangle =\frac{n^{2}}{\sqrt{1+\xi }},  \label{eq:E_n_PT_2}
\end{equation}
that yields the correct linear term (\ref{eq:E_n_PT}) and is singular at $%
\xi =-1$.

\section{Rayleigh-Ritz method}

\label{sec:RRM}

The RRM\cite{P68} is a well known variational procedure that provides
increasingly accurate upper bounds\cite{M33, F25a}. In order to apply this
approach we need a suitable basis set.

The standard Fourier basis set is suitable for the application of the RRM.
However, we found it easier to calculate the necessary matrix elements, by
means of our old-fashioned computer-algebra software, with the
non-orthogonal basis sets
\begin{eqnarray}
(+,+) &:&\;\left\{ \cos ^{2n}\phi ,\;n=0,1,\ldots \right\} ,  \nonumber \\
(+,-) &:&\left\{ \cos ^{2n+1}\phi ,\;n=0,1,\ldots \right\} ,  \nonumber \\
(-,+) &:&\;\left\{ \sin \phi \cos ^{2n+1}\phi ,\;n=0,1,\ldots \right\} ,
\nonumber \\
(-,-) &:&\;\left\{ \sin \phi \cos ^{2n}\phi ,\;n=0,1,\ldots \right\} .
\label{eq:basis}
\end{eqnarray}
We followed a brute-force procedure consisting of obtaining the roots of the
secular determinant $\left| \mathbf{H}-W\mathbf{S}\right| $ where the
elements of the $N\times N$ matrices $\mathbf{H}$ and $\mathbf{S}$ are given
by\cite{P68,F24}
\begin{equation}
H_{ij}=\left\langle \varphi _{i}\right| H\left| \varphi _{j}\right\rangle
,\;S_{ij}=\left\langle \varphi _{i}\right| \left. \varphi _{j}\right\rangle
,\;i,j=0,1,\ldots ,N-1.  \label{eq:H,S_mat_el}
\end{equation}
It is worth noting that the expansion of the eigenfunctions of $H$ in
Fourier series requires the use of the scalar product (\ref{eq:scalar_prod}).

Tables \ref{RRM(+,+)}, \ref{RRM(+,-)}, \ref{RRM(-,-)} and \ref{RRM(-,+)}
show the rate of convergence of the RRM eigenvalues for $\xi =1$ in terms of
the dimension $N$ of the secular determinant\cite{P68,F24}. If we order the
eigenvalues $E_{n}$ in such a way that $E_{n+1}>E_{n}$, we appreciate that
the states with $n>0$ are two-fold degenerate within the accuracy of present
calculation (10 digits). The two states that share the eigenvalue $E_{2n}$
have symmetries $(+,+)$ and $(-,+)$, while those that share $E_{2n+1}$ have
symmetries $(+,-)$ and $(-,-)$. Besides, those four tables suggest that the
eigenvalues of the model can be written
\begin{equation}
E_{n}=n^{2}E_{1},\;n=0,\pm 1,\pm 2,\ldots ,\;E_{1}>0,  \label{eq:conjecture}
\end{equation}
in analogy to the particle on a circular path. This type of spectrum is
reasonable in the case of the circumference because it exhibits a symmetry
axis $C_{\infty }$ but one does not expect it in a model associated to an
ellipse with symmetry axes $C_{2}$. It is clear that there is additional
symmetry in this problem.

The additional symmetry is given by $[H,A]=0$. If $\psi $ is an
eigenfunction of $H$ with eigenvalue $E\neq 0$, then $HA\psi =AH\psi =EA\psi
$. The operator $A$ does not affect the periodicity of $\psi $ but changes
its symmetry; for this reason $\psi $ and $A\psi $ are linearly independent.
We conclude that every eigenvalue $E\neq 0$ is at least two-fold degenerate.
This argument explains the degeneracy revealed by the numerical calculation
discussed above. More precisely, it predicts correctly that the degenerate
states should have different parity as we have already found.

\begin{figure}[tbp]
\begin{center}
\includegraphics[width=9cm]{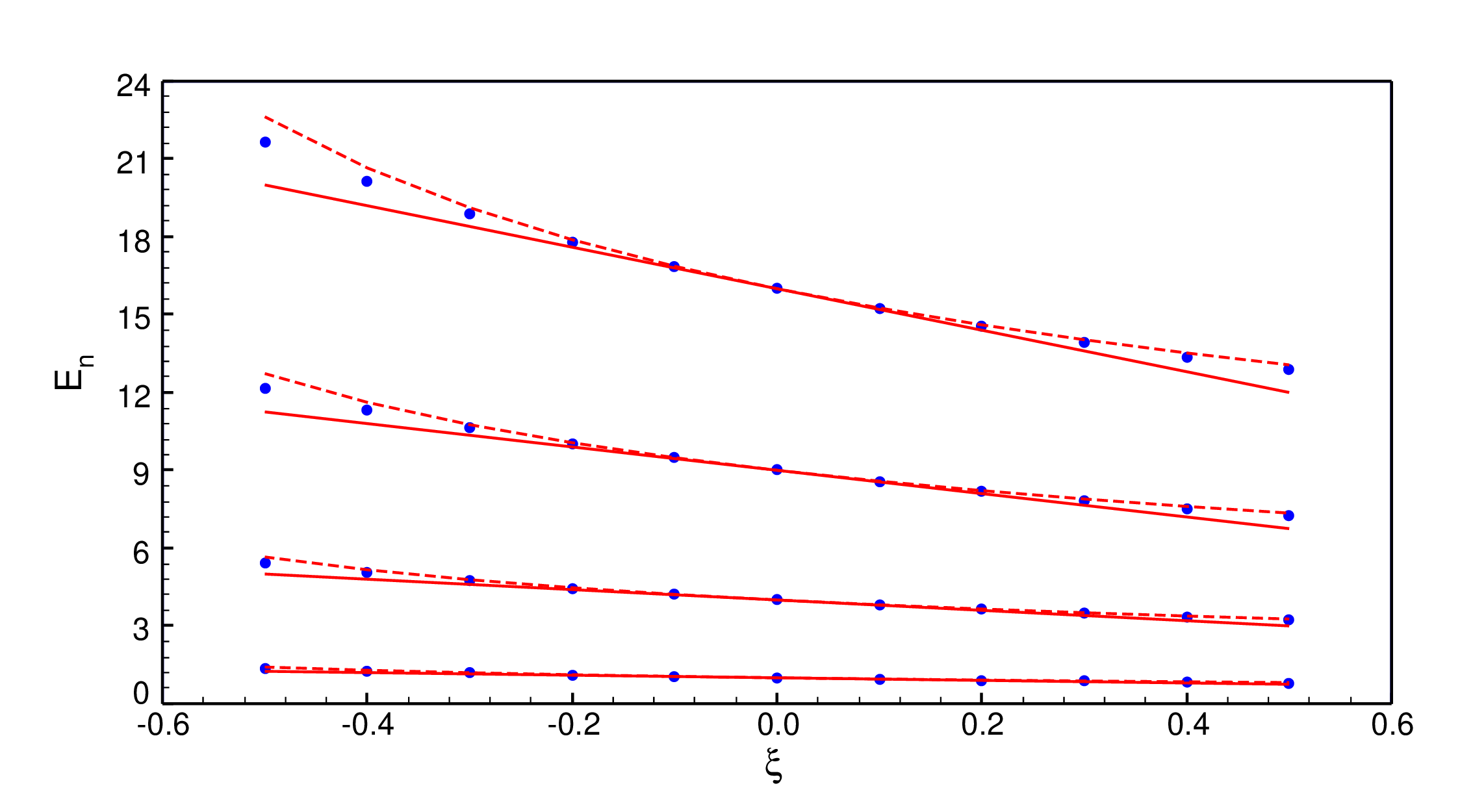}
\end{center}
\caption{RRM (blue circles), PT (solid red line) and improved PT (dashed red
line) eigenvalues with $n=1,2,3,4$ of model (\ref{eq:Schro}) }
\label{Fig:PT}
\end{figure}

Figure~\ref{Fig:PT} shows the RRM eigenvalues and the PT ones given by
equations (\ref{eq:E_n_PT}) and (\ref{eq:E_n_PT_2}) with $n=1,2,3,4$ for $%
-0.5\leq \xi \leq 0.5$. We appreciate that the accuracy of PT decreases with
$n$ and that equation (\ref{eq:E_n_PT_2}) provides a noticeably improvement
over (\ref{eq:E_n_PT}).

We can obtain perturbation corrections of greater order by means of a
straightforward procedure. We substitute the perturbation series (\ref{eq:PT
series}) and the Taylor expansion of $g^{-1/2}$ about $\xi =0$ into the
secular determinant and solve for the perturbation coefficients $E_{n}^{(j)}$%
. In this way we obtain
\begin{eqnarray}
E_{1} &=&1-\frac{1}{2}\xi +\frac{9}{32}\xi ^{2}-\frac{11}{64}\xi ^{3}+\frac{%
917}{8192}\xi ^{4}+\mathcal{O}\left( \xi ^{5}\right) ,  \nonumber \\
E_{2} &=&4-2\xi +\frac{9}{8}\xi ^{2}-\frac{11}{16}\xi ^{3}+\frac{917}{2048}%
\xi ^{4}+\mathcal{O}\left( \xi ^{5}\right) ,  \nonumber \\
E_{3} &=&9-\frac{9}{2}\xi +\frac{81}{32}\xi ^{2}-\frac{99}{64}\xi ^{3}+\frac{%
8253}{8192}\xi ^{4}+\mathcal{O}\left( \xi ^{5}\right) ,  \nonumber \\
E_{4} &=&16-8\xi +\frac{9}{2}\xi ^{2}-\frac{11}{4}\xi ^{3}+\frac{917}{512}%
\xi ^{4}+\mathcal{O}\left( \xi ^{5}\right) ,  \nonumber \\
E_{5} &=&25-\frac{25}{2}\xi +\frac{225}{32}\xi ^{2}-\frac{275}{64}\xi ^{3}+%
\frac{22925}{8192}\xi ^{4}+\mathcal{O}\left( \xi ^{5}\right) ,  \nonumber \\
E_{6} &=&36-18\xi +\frac{81}{8}\xi ^{2}-\frac{99}{16}\xi ^{3}+\frac{8253}{%
2048}\xi ^{4}+\mathcal{O}\left( \xi ^{5}\right) ,  \label{eq:PT_series_mod_1}
\end{eqnarray}
which clearly confirm the conjecture (\ref{eq:conjecture}).

\section{Exact solutions}

\label{sec:exact_sol}

The results shown in sections \ref{sec:PT} and \ref{sec:RRM} are somewhat
irrelevant because the Scr\"{o}dinger equation for the Hamiltonian operator (%
\ref{eq:H_dim}) is exactly solvable. In fact, if we carry out the
transformation $\phi \rightarrow s(\phi )$ and take into account that $\frac{%
d}{d\phi }=s^{\prime }\frac{d}{ds}$ we realize that $\frac{1}{\sqrt{g}}\frac{%
d}{d\phi }=\frac{d}{ds}$ when $s^{\prime }=\sqrt{g}$. We thus conclude that
\begin{equation}
H=-\frac{d^{2}}{ds^{2}},\;s(\phi )=\int_{0}^{\phi }\sqrt{g}d\phi .
\label{eq:H_dim_s}
\end{equation}
The interval $0\leq \phi \leq 2\pi $ maps into $0\leq s\leq L(\xi )$, where
\begin{equation}
L(\xi )=s(2\pi )=\int_{0}^{2\pi }\sqrt{1+\xi \cos ^{2}\phi }d\phi .
\label{eq:L(xi)}
\end{equation}
The exact eigenfunctions and eigenvalues are therefore given by
\begin{equation}
\psi _{n}(s)=\frac{e^{i2\pi ns/L}}{\sqrt{L}},\;E_{n}=\frac{4\pi ^{2}n^{2}}{%
L^{2}},\;n=0,\pm 1,\pm 2,\ldots .  \label{eq:psi_n,E_n,exact}
\end{equation}
One can easily verify that this expression for the eigenvalues yields all
the RRM and PT results obtained in section~\ref{sec:RRM}.

\section{Alternative model}

\label{sec:alternative_mod}

The model discussed in section~\ref{sec:model} was written as $H=-AA$. We
can derive an Hermitian variant by simply writing $H=A^{\dagger }A$; that is
to say
\begin{equation}
H=-\frac{d}{d\phi }g^{-1}\frac{d}{d\phi },  \label{eq:H_2}
\end{equation}
that is Hermitian with the scalar product (\ref{eq:scalar_prod}). Note that
\begin{equation}
H(\ref{eq:Schro})-H(\ref{eq:H_2})=-\frac{g^{\prime }}{2g^{2}}\frac{d}{d\phi }%
.
\end{equation}

We carry out the same RRM calculation as in the preceding model. Table~\ref
{tab:GPTM} shows the lowest eigenvalues for each of the symmetries discussed
above and $\xi =1$. We appreciate that the degeneracy is broken but the
magnitude of the splitting decreases with $n$. It is worth mentioning that
in this case we also have the exact solution $E_{0}=0$, $\psi _{0}(\phi )=%
\frac{1}{\sqrt{2\pi }}$.

By means of the perturbation expansion based on the secular determinant
already described above we obtain
\begin{eqnarray}
E_{1}(-,-) &=&1-\frac{3}{4}\xi +\frac{71}{128}\xi ^{2}-\frac{1655}{4096}\xi
^{3}+\frac{113807}{393216}\xi ^{4}+\mathcal{O}\left( \xi ^{5}\right) ,
\nonumber \\
E_{1}(+,-) &=&1-\frac{1}{4}\xi +\frac{7}{128}\xi ^{2}-\frac{41}{4096}\xi
^{3}+\frac{527}{393216}\xi ^{4}+\mathcal{O}\left( \xi ^{5}\right) ,
\nonumber \\
E_{2}(+,+) &=&4-2\xi +\frac{11}{12}\xi ^{2}-\frac{3}{8}\xi ^{3}+\frac{1781}{%
13824}\xi ^{4}+\mathcal{O}\left( \xi ^{5}\right) ,  \nonumber \\
E_{2}(-,+) &=&4-2\xi +\frac{17}{12}\xi ^{2}-\frac{9}{8}\xi ^{3}+\frac{12533}{%
13824}\xi ^{4}+\mathcal{O}\left( \xi ^{5}\right) ,  \nonumber \\
E_{3}(-,-) &=&9-\frac{9}{2}\xi +\frac{657}{256}\xi ^{2}-\frac{7281}{4096}\xi
^{3}+\frac{7505613}{5242880}\xi ^{4}+\mathcal{O}\left( \xi ^{5}\right) ,
\nonumber \\
E_{3}(+,-) &=&9-\frac{9}{2}\xi +\frac{657}{256}\xi ^{2}-\frac{5823}{4096}\xi
^{3}+\frac{3773133}{5242880}\xi ^{4}+\mathcal{O}\left( \xi ^{5}\right) ,
\nonumber \\
E_{4}(+,+) &=&16-8\xi +\frac{68}{15}\xi ^{2}-\frac{14}{5}\xi ^{3}+\frac{5878%
}{3375}\xi ^{4}+\mathcal{O}\left( \xi ^{5}\right) ,  \nonumber \\
E_{4}(-,+) &=&16-8\xi +\frac{68}{15}\xi ^{2}-\frac{14}{5}\xi ^{3}+\frac{6628%
}{3375}\xi ^{4}+\mathcal{O}\left( \xi ^{5}\right) ,  \nonumber \\
E_{5}(-,-) &=&25-\frac{25}{2}\xi +\frac{5425}{768}\xi ^{2}-\frac{2225}{512}%
\xi ^{3}+\frac{566374475}{198180864}\xi ^{4}+\mathcal{O}\left( \xi
^{5}\right) ,  \nonumber \\
E_{5}(+,-) &=&25-\frac{25}{2}\xi +\frac{5425}{768}\xi ^{2}-\frac{2225}{512}%
\xi ^{3}+\frac{566374475}{198180864}\xi ^{4}+\mathcal{O}\left( \xi
^{5}\right) .  \label{eq:PT_mod_2}
\end{eqnarray}
These analytical expressions suggest that the splitting of the nth level
takes place at perturbation order $n$.

\begin{figure}[tbp]
\begin{center}
\includegraphics[width=9cm]{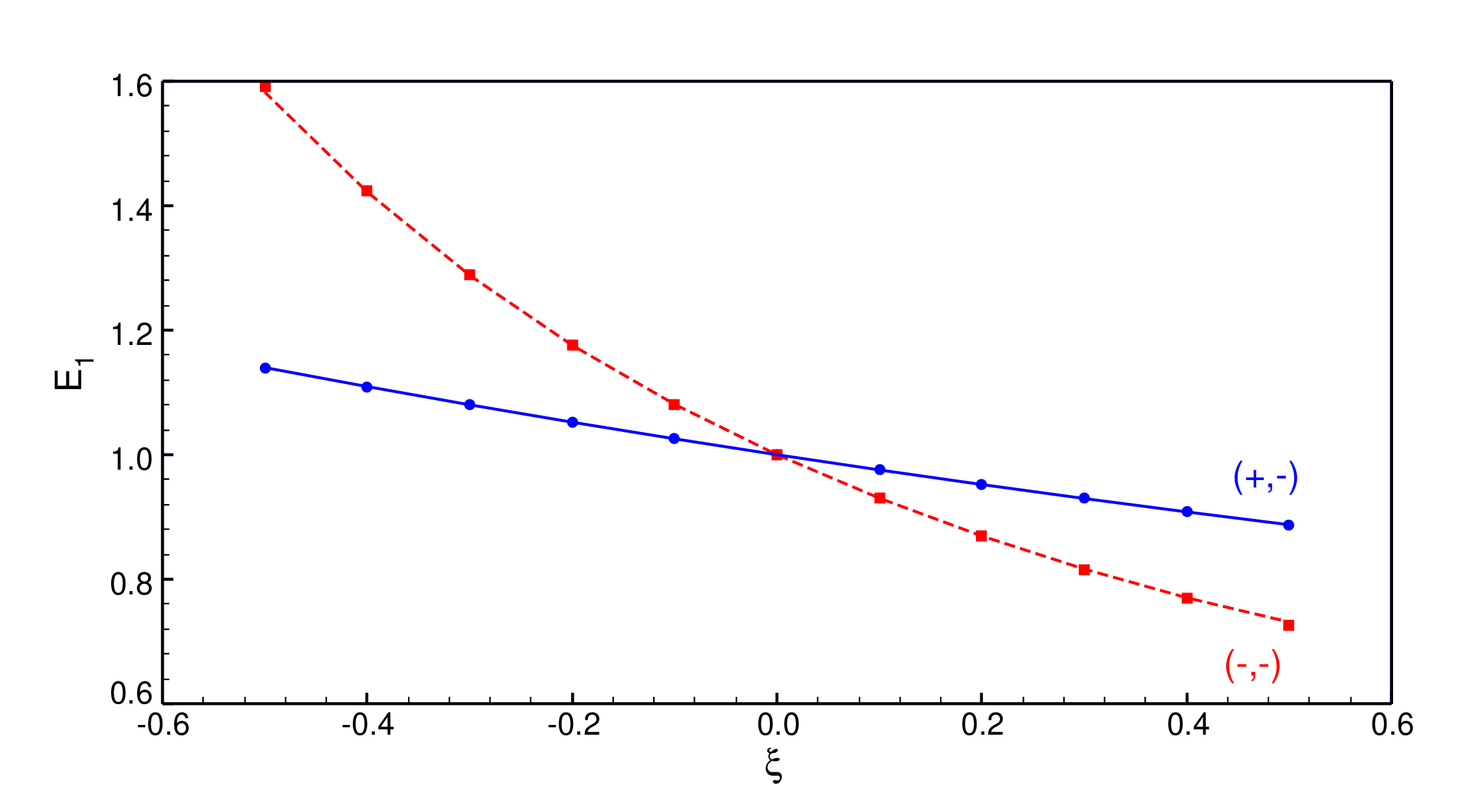}
\end{center}
\caption{RRM eigenvalues (blue circles and red squares) and PT series (blue
solid and red dashed lines) for the first and second excited states of model
(\ref{eq:H_2})}
\label{Fig:Eeo1}
\end{figure}

Figure~\ref{Fig:Eeo1} shows accurate RRM results and PT ones for $E_{1}(-,-)$
and $E_{1}(+,-)$ in a scale that clearly reveals the splitting of the energy
level with $n=1$. We appreciate that the perturbation expansions are
reasonably accurate in this range of values of $\xi $. The first-order
perturbation expression (\ref{eq:E_n_PT_2}) is not suitable for the
description of these two levels with the required detail.

\section{Conclusions}

\label{sec:conclusions}

We have explored two Hamiltonian operators derived from the model of a
quantum-mechanical particle on an elliptical path. The first one is
non-Hermitian but equivalent to an Hermitian operator. For this reason its
eigenvalues are real. The most relevant feature of this quantum-mechanical
model is that it exhibits the same two-fold degeneracy as in the case $\xi
=0 $ (particle on a circular path). Both accurate numerical results and
perturbation theory suggest that the eigenvalues follow the expression shown
in equation (\ref{eq:conjecture}). In addition to it, there is an exact
solution given by a constant eigenfunction and $E_{0}=0$. We can separate
the states of the system into four symmetry species which facilitates the
calculation and the analysis of the problem. An additional symmetry explains
the two-fold degeneracy of the model. The Schr\"{o}dinger equation for this
model is exactly solvable and the exact eigenvalues enable us to confirm the
approximate results provided by the RRM and PT.

The second example is an Hermitian modification of the previous Hamiltonian
operator that exhibits the same type of point-group symmetry but lacks of
the additional symmetry. For this reason the two-fold degeneracy at $\xi =0$
is broken when $\xi \neq 0$. Present low order perturbation expansions
suggests that the splitting of the nth excited level takes place at the nth
perturbation order.

\begin{table}[tbp]
\caption{$(+,+)$ RRM eigenvalues of model (\ref{eq:Schro}) for $\xi=1$}
\label{RRM(+,+)}
\begin{center}
\begin{tabular}{rclll}
$N$ & \multicolumn{1}{c}{$n=0$} & \multicolumn{1}{c}{$n=2$} &
\multicolumn{1}{c}{$n=4$} & \multicolumn{1}{c}{$n=6$} \\
5 & 0 & 2.705129367 & 10.82054064 & 24.39391541 \\
6 & 0 & 2.705129365 & 10.82051747 & 24.34655624 \\
7 & 0 & 2.705129365 & 10.82051746 & 24.34616541 \\
8 & 0 & 2.705129365 & 10.82051746 & 24.34616429 \\
9 & 0 & 2.705129365 & 10.82051746 & 24.34616429 \\
10 & 0 & 2.705129365 & 10.82051746 & 24.34616429
\end{tabular}
\end{center}
\end{table}

\begin{table}[tbp]
\caption{$(-,+)$ RRM eigenvalues of model (\ref{eq:Schro}) for $\xi=1$}
\label{RRM(-,+)}
\begin{center}
\begin{tabular}{rclll}
$N$ & \multicolumn{1}{c}{$n=2$} & \multicolumn{1}{c}{$n=4$} &
\multicolumn{1}{c}{$n=6$} & \multicolumn{1}{c}{$n=8$} \\
5 & 2.705129365 & 10.82051747 & 24.34655624 & 43.45817399 \\
6 & 2.705129365 & 10.82051746 & 24.34616541 & 43.28492974 \\
7 & 2.705129365 & 10.82051746 & 24.34616429 & 43.28208765 \\
8 & 2.705129365 & 10.82051746 & 24.34616429 & 43.2820699 \\
9 & 2.705129365 & 10.82051746 & 24.34616429 & 43.28206985 \\
10 & 2.705129365 & 10.82051746, & 24.34616429, & 43.28206985
\end{tabular}
\end{center}
\end{table}

\begin{table}[tbp]
\caption{$(+,-)$ RRM eigenvalues of model (\ref{eq:Schro}) for $\xi=1$}
\label{RRM(+,-)}
\begin{center}
\begin{tabular}{rclll}
$N$ & \multicolumn{1}{c}{$n=1$} & \multicolumn{1}{c}{$n=3$} &
\multicolumn{1}{c}{$n=5$} & \multicolumn{1}{c}{$n=7$} \\
5 & 0.6762823414 & 6.086541072 & 16.9071682 & 33.23425983 \\
6 & 0.6762823414 & 6.086541072 & 16.90705871 & 33.13897689 \\
7 & 0.6762823414 & 6.086541072 & 16.90705853 & 33.13783974 \\
8 & 0.6762823414 & 6.086541072 & 16.90705853 & 33.13783474 \\
9 & 0.6762823414 & 6.086541072 & 16.90705853 & 33.13783472 \\
10 & 0.6762823414 & 6.086541072 & 16.90705853 & 33.13783472
\end{tabular}
\end{center}
\end{table}

\begin{table}[tbp]
\caption{$(-,-)$ RRM eigenvalues of model (\ref{eq:Schro}) for $\xi=1$}
\label{RRM(-,-)}
\begin{center}
\begin{tabular}{rclll}
$N$ & \multicolumn{1}{c}{$n=1$} & \multicolumn{1}{c}{$n=3$} &
\multicolumn{1}{c}{$n=5$} & \multicolumn{1}{c}{$n=7$} \\
5 & 0.6762823414 & 6.086541072 & 16.9071682 & 33.23425983 \\
6 & 0.6762823414 & 6.086541072 & 16.90705871 & 33.13897689 \\
7 & 0.6762823414 & 6.086541072 & 16.90705853 & 33.13783974 \\
8 & 0.6762823414 & 6.086541072 & 16.90705853 & 33.13783474 \\
9 & 0.6762823414 & 6.086541072 & 16.90705853 & 33.13783472 \\
10 & 0.6762823414, & 6.086541072, & 16.90705853, & 33.13783472
\end{tabular}
\end{center}
\end{table}

\begin{table}[tbp]
\caption{Lowest eigenvalues of model (\ref{eq:H_2}) for $\xi=1$}
\label{tab:GPTM}
\begin{center}
\begin{tabular}{D{.}{.}{11}D{.}{.}{11}D{.}{.}{11}D{.}{.}{11}}

\multicolumn{1}{c}{$(+,+)$}        &   \multicolumn{1}{c}{$(-,+)$}
&      \multicolumn{1}{c}{$(+,-)$}        &
\multicolumn{1}{c}{$(-,-)$}
\\
0             &              &                    &               \\
2.642467139  &  2.79431927   &     0.7959412608  &    0.5700037793 \\
10.81697747  &  10.84750548  &     6.135514729   &    6.062735007  \\
24.35498746  &  24.35945236  &     16.92430649   &    16.91237359  \\
43.29263030  &  43.29320664  &     33.14957349   &    33.1479519   \\

\end{tabular}
\end{center}
\end{table}

\end{document}